\documentstyle[12pt]{article}                                               
\newcommand{\simlt}  {\raisebox{-.6ex}{$\stackrel{\textstyle <}{\sim}$}}

\begin{document}                                                                
\begin{flushright}
RAL-TR-95-078 \\                                                             
20 December 1995 \\                                                               
\end{flushright}                                                               
\vspace{0 mm}                                                                   
\begin{center}
{\Large  
No Mikheyev-Smirnov-Wolfenstein Effect
in Maximal Mixing}            
\end{center}
\vspace{5mm}
\begin{center}                      
{P. F. Harrison\\                   
Physics Department, Queen Mary and Westfield College\\ 
Mile End Rd. London E1 4NS. UK \footnotemark[1]}     
\end{center}                        
\begin{center}
{and}                               
\end{center}                        
\begin{center}                      
{D. H. Perkins\\                      
Nuclear Physics Laboratory, University of Oxford\\
Keble Road, Oxford OX1 3RH. UK \footnotemark[2]} 
\end{center}
\begin{center}                      
{and}                               
\end{center}
\begin{center}                      
{W. G. Scott\\                      
Rutherford Appleton Laboratory\\    
Chilton, Didcot, Oxon OX11 0QX. UK \footnotemark[3]} 
\end{center}
\vspace{1mm}
\begin{abstract}
\baselineskip 0.6cm
We investigate the possible influence of the MSW effect
on the expectations for the solar neutrino experiments
in the maximal mixing scenario suggested by the
atmospheric neutrino data.
A direct numerical calculation 
of matter induced effects in the Sun shows that the
naive vacuum predictions are left completely undisturbed
in the particular case of maximal mixing,
so that the MSW effect turns out to be unobservable.
We give a qualitative explanation of this result.
\end{abstract}
\begin{center}
{\em To be published in Physics Letters B}
\end{center}
\footnotetext[1]{E-mail:P.F.Harrison@QMW.AC.UK}
\footnotetext[2]{E-mail:D.Perkins1@OX.AC.UK}
\footnotetext[3]{E-mail:W.G.Scott@RL.AC.UK}
\newpage 
\baselineskip 0.6cm
There is no doubt that the famous
Mikheyev-Smirnov-Wolfenstein (MSW) mechanism 
\cite{WOLF} \cite{MS} \cite{BETHE}
continues to provide an elegant and viable
explanation for the existing solar neutrino data 
\cite{HOME} \cite{KAMS} \cite{SAGE} \cite{GALLEX}.
The preferred MSW fit requires one neutrino
mass-squared difference $\sim 10^{-5}$ eV$^2$
and one small mixing angle $\sim 10^{-2}$ radians.
On the other hand large mixing,
and in particular maximal mixing \cite{NUSS}
is not completely ruled out by the existing solar data
and is in fact actively suggested by independent data relating to
atmospheric neutrinos \cite{KAMA} \cite{IMB} \cite{SOUDAN} \cite{MACRO}.
The atmospheric neutrino data require a larger neutrino mass-squared
difference $\sim 10^{-2}$ eV$^2$.
In this paper we focus attention
on the maximal mixing scenario \cite{HPS} \cite{KIM}
suggested by the atmospheric data and proceed to investigate
the possible influence of the MSW effect
on the expectations for the solar neutrino experiments in that case.
Our main results are based on a direct numerical
calculation of matter induced effects in the Sun.
We find that, in the maximal mixing scenario,
matter induced effects turn out to be essentially unobservable,
with the naive vacuum predictions
left completely undisturbed,
in the specific case of maximal mixing.
 
The MSW effect has its origin in the interaction
of the solar neutrinos with the matter in the Sun.
In particular in the presence of matter the neutrino mass matrix
is modified by the forward scattering
of electron-neutrinos from electrons via the weak charged current.
In a weak basis which diagonalises
the mass matrix of the charged leptons,
the $3 \times 3$ vacuum propagation matrix $m^2/2E$
is replaced by a matter propagation matrix
$m^2/2E+$diag$(\sqrt{2}GN_{e},0,0)$,
where $mm^{\dagger} \equiv m^2$ is 
the hermitian-square of the vacuum neutrino mass matrix,
$E$ is the neutrino energy, $G$ is the Fermi constant,
and $N_{e}$ is the (position dependent)
number density of electrons in the Sun.

We calculate the MSW effect
numerically for arbitrary $3 \times 3$ vacuum mixing.
To specify the vacuum mixing we take over
the standard parameterisation \cite{PDG}
of the quark mixing matrix,
so that, for example, threefold maximal mixing \cite{HPS}
is reproduced by setting $\theta_{12}=\theta_{23}=\pi/4$,
$\theta_{13}=\sin^{-1}(1/\sqrt{3})$ and $\delta = \pi/2$.
For given values of the input mixing parameters,
we first construct a vacuum neutrino mass matrix
(in the above basis)
as a function of the two independent neutrino mass-squared differences
$\Delta m^2$ and $\Delta m'^2$ ($\Delta m^2 \ge \Delta m'^2$ \cite{HPS}).
To account for matter effects
we divide the propagation path longitudinally
into thin slices of (variable) thickness $\Delta$.
For a given slice we calculate
the matrix of transition amplitudes
just as in the vacuum case ($A=\exp(-im^2\Delta/2E)$),
but using the matter propagation matrix calculated
assuming a constant density over the slice.
The overall matrix of transition amplitudes 
is given by the ordered product
of those for the individual slices, and the final
electron-neutrino survival probability P($e \rightarrow e$)
is averaged over neutrino energies
for comparison with experiment.

Our calculation is `exact'
(at least in the limit $\Delta \rightarrow 0$)
in the sense that it does not rely on any particular
physical approximation relating to the MSW effect itself
(eg.\ `adiabatic approximation' etc.\ ).
In practice, for the results presented below,
we consider only radially directed
propagation paths starting from the center of the Sun
(the detailed production profile is unimportant here)
and we average over a uniform distribution
of neutrino energies with a width of $\pm 25$\% \cite{HPS}.
The number density of electrons in the Sun
as a function of radius $R$ is parameterised \cite{BAH} by
$N_{e} = N_{A} \exp (5.50-\sqrt{(10.54R/R_{\odot})^2+0.89^2})$,
where $N_{A} = 6.02 \times 10^{23}$ cm$^{-3}$
and $R_{\odot} = 0.7 \times 10^6$ km is the solar radius.
For given $\Delta m^2$ and $\Delta m'^2$,
typically $\sim 1000$ propagation slices 
and $\sim 2500$ energy samplings
proved sufficient to produce a robust result.
High precision (128-bit) arithmetic
was found to be neccessary for carrying out
the matrix manipulations. 

We first present results for
the case of $2 \times 2$ mixing,
ie.\ for a simplified scenario in which one of the three generations
is completely decoupled in the mixing matrix
and may in effect be forgotten.
In the $2 \times 2$ case
the mixing is completely specified
by a single mass-squared difference $\Delta m^2$,
and by a single mixing angle $\theta$.
Figure~1 shows our results for $2 \times 2$ mixing.
Figure~1a is for vacuum mixing only,
ie.\ with matter induced effects neglected,
and Figure~1b shows our results
with matter induced effects taken into account.
We plot the expected electron survival probability
$P(e \rightarrow e)$ for solar neutrinos, as measured on Earth,
as a function of $\Delta m^2/E$,
for various values of $\sin \theta$ as indicated.
For $\Delta m^2/E$ $\simlt$ $10^{-12}$ eV$^2$/MeV,
the oscillation length is longer than the distance
from the Sun to the Earth and $P(e \rightarrow e)=1$. 
In the case of vacuum mixing
the biggest suppression (a factor of $1/2$)
occurs in the case of twofold maximal mixing
($\sin \theta \equiv 1/\sqrt{2}$).
With matter effects included,
for $\sin \theta < 1/\sqrt{2}$
(eg.\ for $\sin \theta =0.5$, see Figure~1b) 
we reproduce the familiar MSW `bathtub' \cite{RJNP} suppression,
extending over the range
$\Delta m^2/E = 10^{-8}-10^{-5}$ eV$^2$/MeV.
For $\sin \theta > 1/\sqrt{2}$
(eg.\ for $\sin \theta =0.9$, see Figure~1b)
we have an inverted bathtub, ie.\ an MSW enhancement.
The MSW enhancement
occurs when the {\em lighter} charged lepton,
the electron, couples preferentially to the {\em heavier} neutrino
(the rows and columns of the mixing matrix are
ordered in increasing mass).
In the case of twofold maximal mixing
there is neither a suppression nor an enhancement.
The solid curve in Figure~1b
is identical to the corresponding curve in Figure~1a.
We conclude that in the $2 \times 2$ context
the MSW effect is unobservable 
in the particular case of maximal mixing.

Our results for $3 \times 3$ mixing are shown in Figure~2. 
For the $3 \times 3$ case we compute the survival probability
$P(e \rightarrow e)$
as a function of the smaller independent mass-squared difference
$\Delta m'^2$, with the larger mass-squared difference $\Delta m^2$ fixed
by the atmospheric neutrino data ($\Delta m^2 \equiv 10^{-2}$ eV$^2$).
The MSW effect is now governed by $|U_{2e}|$, the magnitude
of the element of the mixing matrix linking the electron
with the second lightest neutrino mass eigenstate.
In the standard parameterisation \cite{PDG}
$U_{2e} = \cos \theta_{13} \sin \theta_{12}$.
For the results presented in Figure~2
we vary $|U_{2e}|$ by varying $\theta_{12}$,
keeping $\theta_{23}$, $\theta_{13}$ and $\delta$ fixed.
For $\Delta m'^2/E$ $\simlt$ $10^{-12}$ eV$^2$/MeV
we have $P(e \rightarrow e) = 5/9$.
Figure~2a is for vacuum mixing only
and Figure~2b shows our results
with matter effects included.
In the case of vacuum mixing
the biggest suppression (a factor of 1/3 for $3 \times 3$ mixing) 
occurs in the case of threefold maximal mixing.
With matter effects included, the MSW effect leads to a 
suppression or an enhancement, in general, 
depending on the value of $\theta_{12}$,
as shown by the broken curves in Figure~2b.
The MSW effect has no influence at all, however,
in the particular case of threefold maximal mixing
({cf.\ }the solid curves in Figure~2b and Figure~2a)
mirroring exactly our results
for the $2 \times 2$ case above.

For completeness it should be said
that if the larger mass-squared difference $\Delta m^2$
were {\em not} constrained by the atmospheric data {\em and} if it fell
in the appropriate range viz.\ $\Delta m^2 = 10^{-8}-10^{-5}$ eV$^2$,
then the MSW effect would become observable
in the case of threefold maximal mixing.
The effect, however
(assuming $\Delta m^2 \gg \Delta m'^2$),
is simply to suppress $P(e \rightarrow e)$ 
by a factor of $1/3$ (instead of $5/9$) over the
range of the bathtub, as shown in Figure~3,
so that again the observable suppression
factors are in general
identical to the case of vacuum mixing.

With a view to obtaining a better understanding of the evident 
`special case' status of maximal mixing
with respect to the MSW effect,
we return, for simplicity, to reconsider the case of $2 \times 2$ mixing,
as a function of $\Delta m^2/E$,
for an arbitrary mixing angle $\theta$, as before.
The effective Hamiltonian for the MSW system
is just the matter propagation matrix above,
which in the $2 \times 2$ case
$\nu_e-\nu_{\mu}$ (say), may be written: 
\begin{equation}
\left( \matrix{ -(\Delta m^2/2E)\cos2\theta+GN_e/\sqrt{2} 
               & (\Delta m^2/2E)\sin2\theta \cr
                 (\Delta m^2/2E)\sin2\theta 
               & (\Delta m^2/2E)\cos2\theta-GN_e/\sqrt{2} } \right).
\end{equation}
In vacuum, in the small $\theta$ limit,
$\nu_{\mu}$ is the heavy mass eigenstate.
In the high density limit $\nu_e$
is the heavy mass eigenstate.
The familiar near-total MSW suppression 
of the $\nu_e$ flux for small mixing angles
occurs when the matter density profile in the Sun
provides a smooth matching from the $\nu_e$ state
at the point of production,
to a near-$\nu_{\mu}$ state outside the Sun.

We exploit an analogy between the $\nu_e - \nu_{\mu}$
system in the presence of a variable matter density, 
and the behaviour of a spin-$1/2$ dipole
at rest in a time-dependent uniform magnetic field.
Suppose that the dipole
has a negative magnetic moment $-\mu$.
Suppose further that the magnetic field {\bf B} 
seen by the dipole may be decomposed as the vector sum
of a constant (ie.\ time independent) `intrinsic' field {\bf B}$^0$
and a variable (ie.\ time dependent) `external' field {\bf B}$_e$.
If the external field {\bf B}$_e$
is directed along the quantisation axis (the $z$-axis)
while the intrinsic field {\bf B}$^0$
makes an angle $2\theta$ with respect to the negative $z$-axis
and is contained in the $zx$-plane, then
the Hamiltonian for the dipole may be written:
\begin{equation}
\left( \matrix{ -\mu B^0 \cos2\theta + \mu B_e 
              &  \mu B^0 \sin2\theta \cr
                 \mu B^0 \sin2\theta 
               & \mu B^0 \cos2\theta - \mu B_e } \right).
\end{equation}
Comparing Eqs.\ 1 and 2 we see that,
with the correspondence $\mu B^0 \leftrightarrow \Delta m^2/(2E)$
and $\mu B_e \leftrightarrow GN_e/\sqrt{2}$,
the dipole and the MSW system 
have the same Hamiltonian.
The utility of the analogy lies in the fact that
the behaviour of the dipole is readily understood,
since the spin vector {\bf S} for the dipole
satisfies a well known classical equation of motion
({\bf \.{S}}$=-2\mu${\bf S}$\wedge${\bf B}).
The dipole simply precesses around the
instantaneous magnetic field {\bf B},
with instantaneous angular frequency $2\mu B$.

The small angle MSW effect
may now be viewed as the adiabatic reversal of the spin of the dipole,
in response to the slow reversal of the field {\bf B},
as the external field {\bf B}$_e$ decreases to zero.
This is illustrated in Figure~4. 
The resonance condition is satisfied
when {\bf B} is directed horizontally along the $x$-axis,
and the mixing becomes momentarily maximal.
If the vacuum mixing is anyway maximal
the intrinsic field {\bf B}$^0$ is directed
entirely horizontally along the $x$-axis
and no such reversal can occur.
The inverted bathtub seen for large mixing angles
corresponds to the case that {\bf B}$^0$ points upwards.
In the maximal mixing case,
the dipole simply follows {\bf B} smoothly from the vertical
to the horizontal and remains horizontal, yielding 
50\% spin-up ($\nu_e$) and 50\% spin-down ($\nu_{\mu}$),
just as for maximal mixing in vacuum.
The difference is that in the matter case 
the residual oscillations are small, 
while in the vacuum case
the dipole precesses around the $x$-axis,
corresponding to maximal (100\%) oscillations.
It is only because
these oscillations are unresolved
that the matter and the vacuum predictions
turn out to be indistinguishable.
Presumably the behaviour in the $3 \times 3$ case
has some closely related explanation, 
but we have not attempted to consider the $3 \times 3$ case 
in the equivalent level of detail.

To summarise, 
a direct numerical calculation
shows that the MSW effect is unobservable
in the particular case of maximal mixing.
The naive vacuum predictions
are left completely undisturbed in that case.
This result is valid for $2 \times 2$ mixing
for any value of $\Delta m^2$,
and for $3 \times 3$ mixing
for any value of $\Delta m'^2$, 
with $\Delta m^2 \sim 10^{-2}$eV$^2$,
fixed by the atmospheric neutrino data.
Exploiting the analogy between the MSW effect
in $2 \times 2$ mixing and the behaviour
of a spin-1/2 dipole in a time-dependent magnetic field,
we have given a qualitative explanation of this result.
It is true nonetheless that
the small angle MSW solution,
with appropriate choices for the parameters,
gives an excellent fit to the existing solar data,
and is currently (perhaps not unnaturally therefore) 
widely accepted as {\em the} solution to the solar neutrino problem.
While our results do not in any way
undermine the validity of the MSW solution,
they do serve to draw attention to an
interesting and significant exception,
where the MSW mechanism cannot be invoked.
The MSW and maximal mixing scenarios
as solutions to the solar neutrino problem
are, in a very definite and real sense, 
to be seen as mutually exclusive alternatives.

\vspace{5mm}
\noindent {\bf Acknowledgement}

\noindent It is a pleasure to thank Roger Phillips
for a number of useful discussions.

\newpage

\noindent {\bf {\large Figure Captions}}

\vspace{10mm}
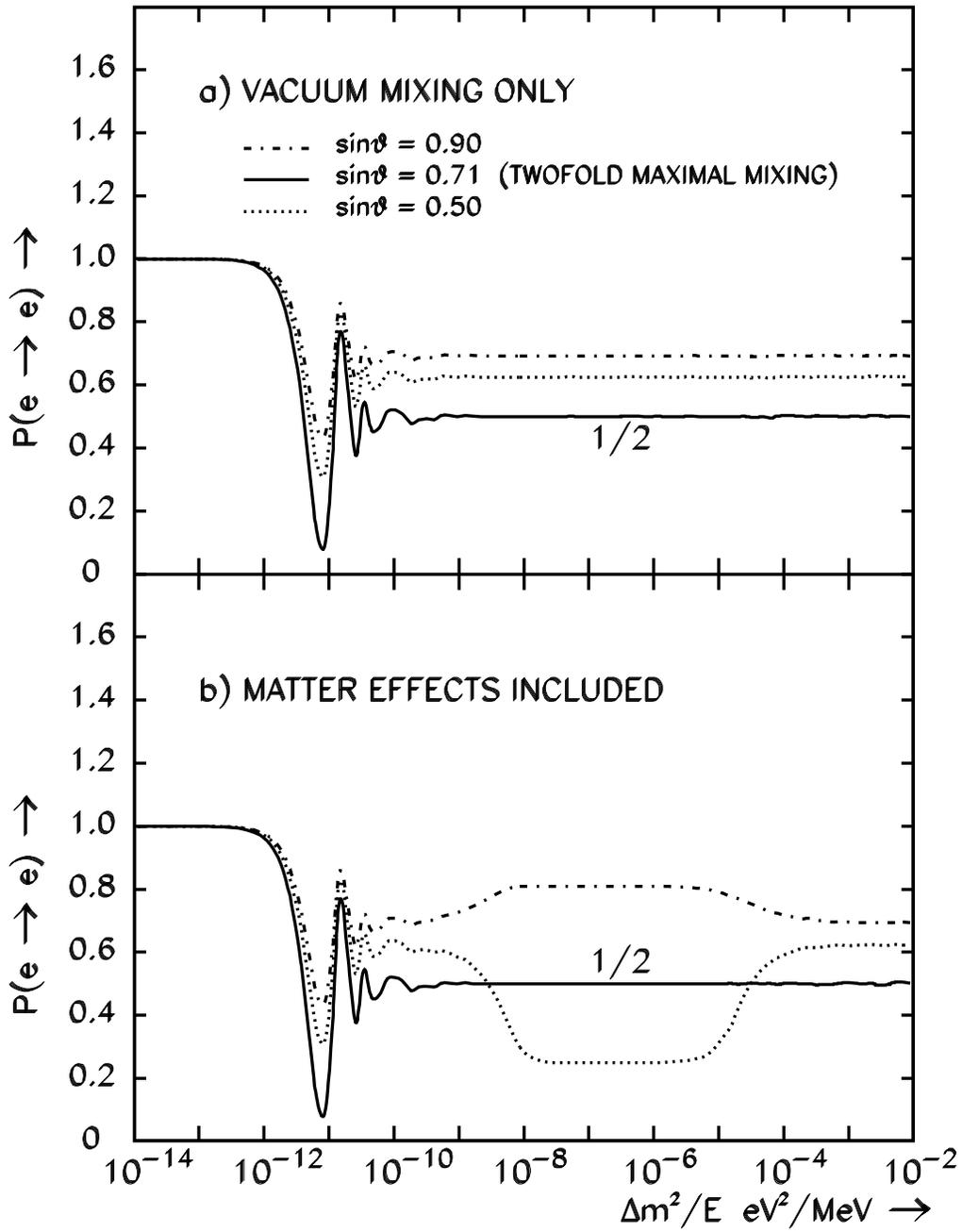
\noindent Figure~1.
The expected electron-neutrino survival probability $P(e \rightarrow e)$, 
for solar neutrinos, as measured on Earth,
for the case of $2 \times 2$ mixing.
The survival probability is plotted as a function of
$\Delta m^2/E$ for various values 
of $\sin \theta$, as indicated, for
a) vacuum mixing and b) accounting for matter induced effects in the Sun.
In the particular case of maximal mixing (solid curves)
the vacuum and matter curves are indistinguishable.

\vspace{10mm}
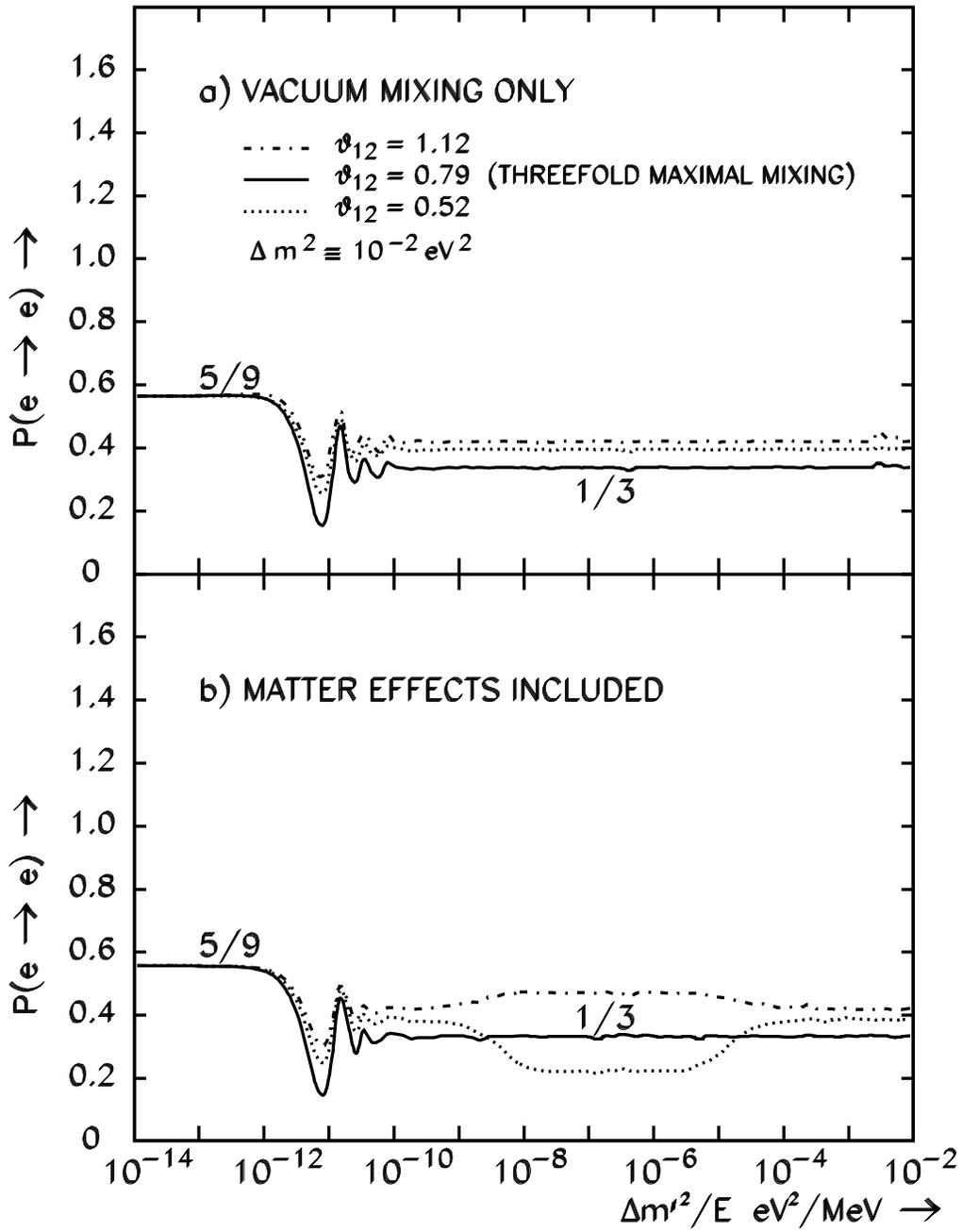
\noindent Figure~2.
The expected electron-neutrino survival probability $P(e \rightarrow e)$, 
for solar neutrinos, as measured on Earth,
for the case of $3 \times 3$ mixing.
The survival probability is plotted as a function of
$\Delta m'^2/E$ (with $\Delta m^2 \equiv 10^{-2}$ eV$^2$)
for various values of $\theta_{12}$, as indicated, for
a) vacuum mixing and b) accounting for matter induced effects in the Sun.
In the particular case of threefold maximal mixing (solid curves)
the vacuum and matter curves are indistinguishable.

\vspace{10mm}
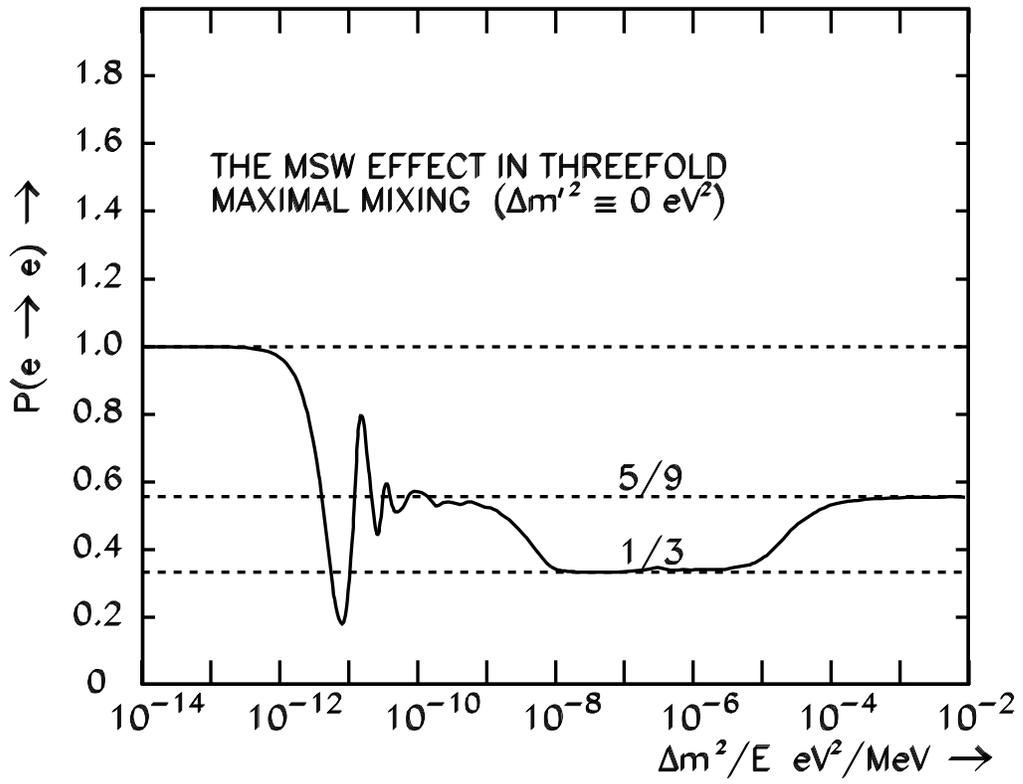
\noindent Figure~3.
If $\Delta m^2$
were not constrained by the atmospheric data, 
the MSW effect would become observable
in threefold maximal mixing 
for $\Delta m^2 = 10^{-8}-10^{-5}$ eV$^2$, as shown
(assuming $\Delta m'^2 \ll \Delta m^2$).
The suppression factor is $1/3$ (instead of $5/9$)
over the range of the `bathtub', however,
so that the observable suppression factors
are anyway identical to those for the vacuum case.

\vspace{10mm}
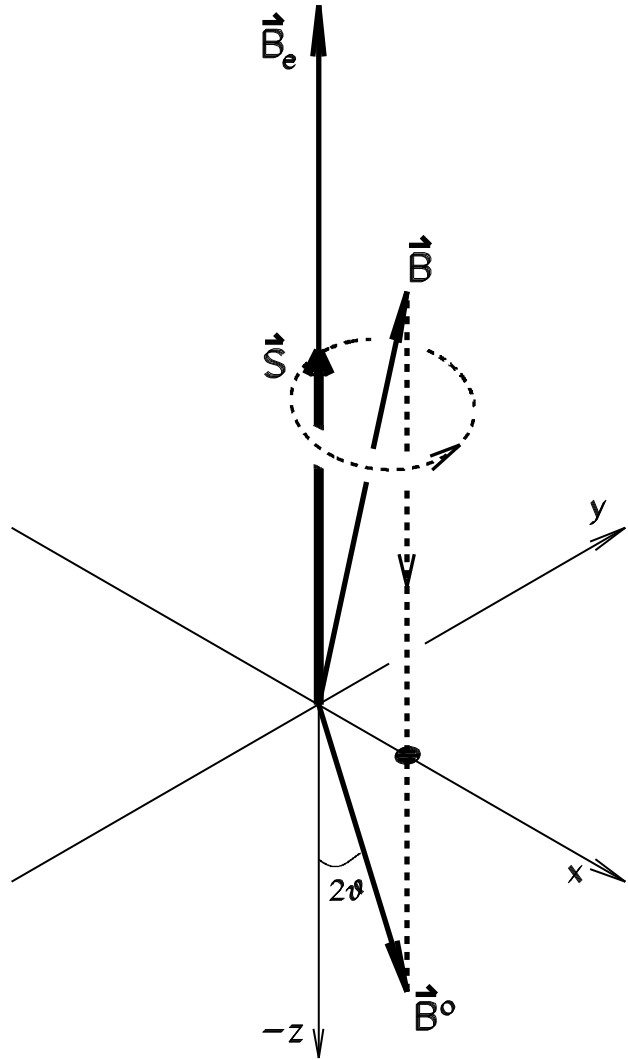
\noindent Figure~4.
The near-total MSW suppression of the $\nu_e$ flux
for small mixing angles
is analogous to the adiabatic reversal of a spin-1/2
dipole in the time dependent magnetic field illustrated.
As the `external' field {\bf B}$_e$ decreases to zero
for fixed `intrinsic' field {\bf B}$^0$ as shown,
the resultant field {\bf B} seen by the dipole reverses.
In the case of maximal mixing the intrinsic field 
{\bf B}$^0$ is directed entirely along the $x$-axis
and no such reversal occurs.

\end{document}